# Global Estimates of Spatially Distributed Surface Energy Fluxes using Thermodynamic Principles

Mayank Gupta[1], Martin Wild[2], and Subimal Ghosh[3,4,*]

[1]Centre for Urban Science and Engineering, Indian Institute of Technology Bombay, Mumbai – 400076, India

[2] ETH Zurich, Institute for Atmospheric and Climate Science, 8001 Zürich, Switzerland

[3]Interdisciplinary Program in Climate Studies, Indian Institute of Technology Bombay, Mumbai – 400076, India

[4] Department of Civil Engineering, Indian Institute of Technology Bombay, Mumbai – 400076, India

[*]Corresponding Author. Email: subimal@iitb.ac.in

## Abstract:

Limited surface observations of turbulent heat fluxes result in incomplete knowledge about the surface energy balance that drives the climate system. Here, we developed a novel, purely physics-based analytical method grounded on the thermodynamic principle of maximum power. The approach derives the turbulent heat flux only from the four inputs of incoming and outgoing radiations at the land surface. The proposed approach does not use any parameterization, unlike the existing surface energy balance models, and hence does not suffer from uncertainty due to the same.We validated our methodology with 102 eddy covariance observation stations around the globe with different land use land covers. Using the satellite observations from CERES at a spatial resolution of $1^0$, we have obtained spatially distributed global analytical estimates of Sensible ($H$), latent heat ($LE$), and land surface heat storage ($\Delta Q_s$) fluxes for the first time. For a global observed land Net radiation ($Q^*$) of 84 Wm$^{-2}$ from the satellite, we found $H$, $LE$ and $\Delta Q_s$ to be 42 Wm$^{-2}$, 40 Wm$^{-2}$ and 2 Wm$^{-2}$, respectively. The theoretical and precise estimates of all surface energy balance components will improve our understanding of surface warming for different land use land covers across the globe.

## Introduction:

The energy flow between the surface and the atmosphere largely determines global and regional changes in the climate system due to variations in the atmospheric conditions and surface state[1,2]. The surface energy balance (SEB) is fundamental to assessing this energy exchange[3]. It disentangles the surface feedback through the competing ecohydrological, biophysical, geophysical processes, and anthropogenic alterations[4–6]. The direct global satellite observations of radiative exchange at the top of the atmosphere combined with the global ocean heat content measurements have established global warming by estimating the changes in Earth's energy imbalance (EEI)[2,7,8]. However, there is incoherent knowledge concerning the distribution of radiative energy at the land surface that is mostly shared by the non-radiative surface energy fluxes and driving factors[3,9]. Primarily, the surface heating due to the absorbed solar radiation dissipates as the energy transfer to the cooler atmosphere by the net exchange of terrestrial longwave radiation and exchange energy as turbulent heat in the form of Sensible and Latent heat flux. Further, the surface retains some part of the energy as a land surface heat storage flux. The different techniques[10–12] to estimate surface fluxes depend highly on site-specific parameters and climatic conditions, creating uncertainty for global assessment[13–15]. Direct *in-situ* observations are collected as point measurements from surface stations; for example, FLUXNET[16,17] is a global network of Eddy covariance (EC) towers but with limited coverage for extrapolating to globally distributed estimates[17,18]. Furthermore, the SEB assessment using the EC towers mostly leaves unexplained residual energy, averaging about 16% of available energy, resulting in a surface energy balance closure problem[19,20]. The regional climate models used for the surface energy balance computation require detailed surface characteristics and high-resolution for better performance[21,22], which makes them computationally intensive[23]. Their outputs in simulating energy exchange are characterized by high uncertainty due to the use of spatially and temporally varying parameters and transfer coefficients, which are not well established[24]. The varied sources of input variables and use of multiple climate models are another sources of uncertainty in SEB simulations[1,25]. Remote sensing has been used as a complementary aid to improve the spatial estimate of turbulent fluxes by estimating land surface temperature, vegetation indices, and surface geometric characteristics for heterogeneous surfaces and different land covers. Still, they need climate or land surface models, where the satellite observations are assimilated[26]. Direct and generalized physics-based estimates of the turbulent fluxes, from satellite observations, independent of surface inputs and parameterization, address the abovementioned limitations. However, such models are yet to be developed to assess the Spatio-temporal land energy feedback to the atmosphere.

Another crucial component of SEB is the land surface heat storage flux. In most of the modeling techniques, it is used as input to estimate turbulent fluxes[27]. More importantly, it determines the inertial heat capacity of the land, develops diurnal variation in local climate, and governs available energy partitioning into sensible and latent heat flux[28]. Along with the soil layer, it comprises canopy heat storage that constitutes heat storage in land cover, biomass, water content, and photosynthesis in the canopy[29–31]. Thus, it shows the potential to explain the role of the variation in land characteristics in climate change. These characteristics are mostly neglected in climate models as they are not easy to measure, need to be parameterized, and are assumed to have insignificant value[19]. However, the net canopy storage can aggregate up to 15 per cent of net radiation for crop sites[30] and 2-6 times more for urban canopies[32,33].

We developed an approach based on principles of thermodynamics to estimate surface fluxes that describe the land-atmosphere as a radiative-convective system. The energy and entropy budget through the first and second law of thermodynamics, respectively, describes directions, constraints, and limits of energy conversion in this system. We reviewed the studies[34–40] based on thermodynamics describing the energy and water exchange through a natural radiative-convective system in equilibrium or steady state. Here, we updated the existing thermodynamic theory and developed a theoretical method to estimate the turbulent fluxes in the SEB directly for the first time. Unlike previous studies, the method does not need *in-situ* measured surface heat storage or its complex parameterization, thus bringing an enlightened perspective on the SEB and its closure for different land covers. We derived an analytical expression (details in Theory and Methods) to estimate the turbulent heat fluxes that does not require high computation requirements and eliminates the highly uncertain parameterization inputs like roughness length. The estimated turbulent fluxes are tested with Eddy covariance observations of the FLUXNET2015[41] database across different ecosystems, and land uses classified by the International Geosphere–Biosphere Programme (IGBP). Further, we showed the potential of the proposed approach in estimating the spatially distributed turbulent (Sensible and latent heat) flux fields by taking inputs from the CERES satellite data and global evaporative stress factor. We validated our approach for several grids, where in-situ FLUXNET2015 EC observations are available

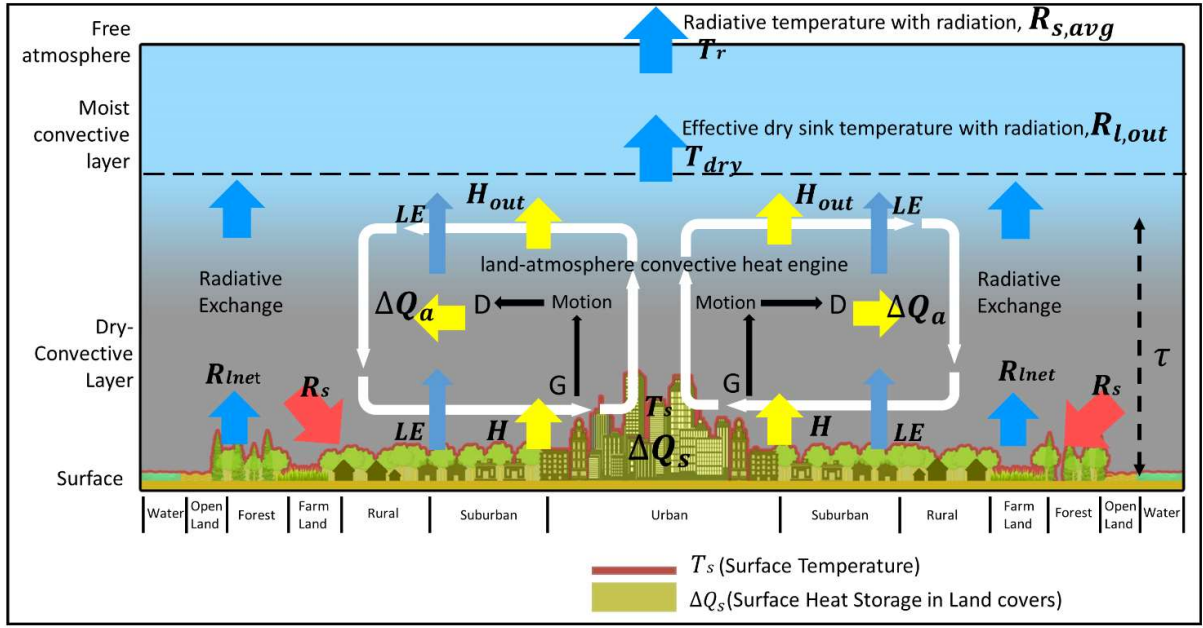

*Figure 1: Schematic diagram of a land-atmospheric convective system using the Thermodynamic theory*

**Theory:**

The Surface energy balance (SEB) at the surface-atmosphere interface is given as (Figure 1):

$$R_s - R_{lnet} = Q_J + \Delta Q_s \qquad (1)$$

Where $R_s$ is the net absorbed shortwave solar radiation flux by the surface, the difference between the incoming shortwave radiation ($K_\downarrow$) and the reflected shortwave radiation ($K_\uparrow$) ($R_s = K_\downarrow - K_\uparrow$). $\Delta Q_s$ is the surface heat storage flux. $R_{lnet}$ is the net longwave radiation flux, given by the difference between the outgoing longwave radiation ($L_\uparrow$) and the incoming longwave radiation ($L_\downarrow$) ($R_{lnet} = L_\uparrow - L_\downarrow$). $Q_J$ is the turbulent heat flux, the sum of sensible heat flux ($H$) and latent heat flux ($LE$), ($Q_J = H + LE$).

Here, we described the land-atmosphere as a radiative-convective system in a steady-state. The system boundary is comprised of two boundary reservoirs (Figure 1), Surface as a hot reservoir with temperature $T_s$ and free atmosphere as a cold reservoir with radiative temperature $T_r$. The temperature difference between the reservoirs drives the energy transfer from the surface to the atmosphere. The convective process in the atmospheric boundary layer manifests as a heat engine that causes turbulent heat exchange by mechanical motion of heated air parcel. The convective process carries sensible heat ($H$) from the warmer land surface to the cooler atmosphere. In moist conditions, evaporation is critical as it consumes a substantial part of surface energy as latent heat to form water vapour. According to the studies[37–39], the convective process near the land surface establishes such that the water vapour is passively transported as mass exchange with the air parcel in the convective motion until the air parcel saturates and condensation occurs. When the water vapour condenses to form the base of clouds, it releases the latent heat causing the convective motion within the clouds, which ultimately dissipates at the atmospheric radiative temperature ($T_r$). Hence, two different energy transfer processes exist in two layers (Figure 1). The first is the dry convection near the surface with mechanical updraft and downdraft of air parcels that transport sensible heat and passively transport latent heat as water vapour. The other is the moist convection due to the condensation release of latent heat that develops convective motion in the clouds. Figure 1 shows the schematics of the dry and moist convective layers.

The radiative temperature ($T_r$) is assessed based on the diurnal behaviour of the land-atmospheric convective system. We used the approach[40] that described the buffering effect of heat in the land-atmospheric system. The surface gets warmer with respect to the cooler atmosphere during the day with the shortwave heating ($R_s$); however, due to the buffering of this heat into the land surface and the atmosphere, the heat radiates back to the free atmosphere at temperature ($T_r$) during both day and night, averagely. The average longwave radiation to the free atmosphere from the system during both day and night is $R_{s,avg} = \sigma T_r^4$.

Further, the dynamics of the land-atmospheric energy balance are governed by the first and the second laws of thermodynamics. In a thermodynamic system, the positive irreversible entropy describes the irreversible nature and direction of the physical transformation and limits the energy available for mechanical work in a heat engine. In a dry convective system near the surface, the only irreversibility associated is the molecular diffusion of sensible heat from the surface to the adjacent atmospheric layer and within the atmosphere, creating an irreversible entropy of sensible heat diffusion term ($\Delta S_{dif}$). According to the study[35], the magnitude of $\Delta S_{dif}$ is negligible. Hence, the dry convection process works as a perfect heat engine with maximum efficiency, as given by the Carnot limit of maximum power ($G_{carnot}$) and maximum efficiency ($\eta_{carnot}$) (Refer Section A.1 in Methods). The irreversibility in moist convection is associated with entropy due to phase change ($\Delta S_{pc}$) and entropy due to water vapor diffusion ($\Delta S_{dv}$). Both entropies are significant and reduce the ability of moist convection to work at maximum efficiency[35,36]. Hence, they should be considered in a moist convective process. Therefore, we developed our theory that considered only the dry convective system that occurs near the land surface with negligible entropy following a perfect heat engine.

Another important aspect of irreversibility in the convective process is the frictional dissipation of total kinetic energy generated by mechanical motions of atmospheric flows. In a complete system, the dissipation occurs as turbulent dissipation ($D_K$) and precipitation-induced dissipation ($D_P$) that become the parts of the system. $D_K$ is the viscous conversion of mechanical energy of air parcel motion to heat, and $D_P$ is the heat dissipation in microscopic shear zones surrounding hydrometeors. In a dry convective system, we only assume $D_K$ to be associated with sensible heat transport with the dissipation of mechanical energy of air parcel.

The irreversible frictional dissipation of mechanical work associated with the convective motion within the same land-atmospheric system makes the heat engine a Dissipative Heat Engine. The studies[35,42] showed that the converted mechanical energy through frictional dissipation increases the internal energy of the system. The heat from frictional dissipation within the engine could not be used as an additional heat source to the engine in addition to the existing heat source to generate mechanical work as it otherwise violates the first law of thermodynamics[42]. Based on this inference, we derived the power of the dissipative heat engine ($G$) through energy and entropy budget (Refer to Section A.2, Eq. M7-M10). The heat dissipation of mechanical work ($D$) and an additional term that depicts the change in the internal energy of the system ($\Delta U$) act such that $G = D = \Delta U$ in a steady state. In dry convection near the land surface, convective heat flux associated with the heat engine is the sensible heat flux ($J_{in} = H$) The internal energy change ($\Delta U$) represents change in the atmospheric heat storage ($\Delta Q_a$)($\Delta U = \Delta Q_a$)(Figure 1). They are used in the expression for power of dissipative heat engine for dry convection ($G_d$) (Refer Section A.2, Eq. M11-M12).

Further, considering only the dry convective engine near the surface, we define the effective dry sink temperature ($T_{dry}$) at which the remaining sensible heat and water vapour is released into the atmosphere. Below the level of $T_{dry}$ (shown by the X-X line in Figure 1), there is no moist convection taking place[39]. The outgoing radiative flux at $T_{dry}$ is given by $R_{l,out}$. The magnitude of $T_{dry}$ is greater than the $T_r$.

In dry convection, the sensible heat flux is expressed in terms of convective vertical mass flux of air and the temperature difference between the surface and the effective dry sink temperature following the works[37–39]:

$$H = c_p J_m (T_s - T_{dry}) \tag{2}$$

Where $c_p$ is the specific heat capacity of the air, $J_m$ is the convective mass flux exchange of the air parcel.

As the mass of the air parcel transports water vapour from the surface to the cloud base, it is associated with latent heat expressed as follows:

$$LE = c_p J_m (q_s - q_{dry}) \tag{3}$$

Where $q_s$ and $q_{dry}$ are the specific humidity of the surface air and the atmosphere. The above equation of $LE$ is associated with the condition that the air had sufficient time and the continuous availability of water to saturate the dry air that rises in the dry convection near the land[37]. In the case of water limiting conditions, the actual $LE$ is given by $LE_{lim}$ based on the water stress factor ($f_w$) that accounts for water limitation for evaporation such that $LE_{lim} = f_w * LE$. The corresponding sensible heat flux in that case is given by $H_{lim}$. As the convection is dependent on the boundary temperatures, the total convective heat flux ($Q_J$) is same for all conditions such that $Q_J = LE + H = LE_{lim} + H_{lim}$ [43].

The difference in specific humidity can be expressed in terms of the temperature difference and the slope of the saturation pressure curve ($s$) following the study[37].

$$e_{sat,T} = 611 * e^{17.6294 \frac{T-273.16}{T-35.86}} \tag{4}$$

$$s = \frac{de_{sat,T}}{dt} = \frac{\lambda e_{sat,T}}{R_v T^2} \tag{5}$$

Linear approximation is expressed as:

$$(q_s - q_{dry}) = \frac{s}{\gamma}(T_s - T_{dry}) \tag{6}$$

Where $\lambda$ is the latent heat of vaporization (2.5. $10^6$ J kg$^{-1}$ K$^{-1}$) and $R_v$ is the gas constant of water vapour (461 J kg$^{-1}$ K$^{-1}$).

Therefore,

$$LE = \frac{s}{\gamma} H \tag{7}$$

The value of $s$ is based on the temperature of the air after it comes in contact with the surface to get heated up and saturated with water vapour. In a real scenario, the actual temperature of air never reaches the surface temperature $T_s$ within a finite time scale. Hence, $s$ cannot be computed with $T_s$. As $s$ varies exponentially with temperature, using $T_s$ for the computation of $s$ will lead to a very high value. Hence, to start with, we estimate the $s$ at the conservative temperature $T_r$ rather than $T_s$. We compensate for the balance energy later in the convective turbulent energy.

Further, equation (M8) and (M9) in dry convection becomes:

$$\Delta Q_a = H - H_{out} \tag{8}$$

$$\frac{\Delta Q_a}{T_a} = \frac{H}{T_s} - \frac{H_{out}}{T_{dry}} + \frac{D_k}{T_a} + \Delta S_{dif} \tag{9}$$

Where $H_{out}$ is the release of sensible heat flux out of the dry convective heat engine, $D_k$ is the turbulent frictional dissipation of the mechanical work within the engine itself, and $\Delta S_{dif}$ is the entropy due to the diffusion of the sensible heat flux which is negligible.

In the steady-state $G_d = D_k = \Delta Q_a$. We assume the dissipation to be at the surface for near-surface atmospheric convection. Therefore, the power, in this case, is given by:

$$G_d = H \cdot \frac{T_s - T_{dry}}{T_s} = \Delta Q_a \tag{10}$$

A simple linearization for $R_{lnet}$ is adopted from the study[43] (Refer Section A.4, eq. M17-M24) to replace $T_s - T_{dry}$ in eq.10 in terms of heat flux and radiative exchanges (Refer Section A.5, Eq M25-M27) to estimate maximum convective power using expression for $G_d$:

$$G_d = H \cdot \frac{R_s - R_{l,0} - H\left(1 + \frac{S}{\gamma}\right) - \Delta Q_s}{T_s K_d} \tag{11}$$

Most of the earth system processes are effectively explained by the maximum power limit[37–39]. There exists a optimal value of sensible heat flux, $H_{opt}$, at which the convective power is maximum ($G_{d,max}$). The value of $H_{opt}$ at $G_{d,max}$ is derived by $\frac{dG_d}{dH} = 0$.

Solving $\frac{dG_d}{dH} = 0$, to obtain the analytical expression for $H_{opt}$ which is given as:

$$cH_{opt} = \frac{1}{2}(R_s - R_{l,0} - \Delta Q_s) \tag{12}$$

Where, $c = 1 + \frac{S}{\gamma}$

The total turbulent flux in terms of surface heat storage based on the maximum convective power is given by the eq. 12.

We use the radiation components, $K_\downarrow, K_\uparrow, L_\uparrow$ and $L_\downarrow$, as input variables to estimate turbulent flux.

The surface net-all wave radiation ($Q^*$) is given by:

$$Q^* = R_s - R_{l,net} \tag{13}$$

$$Q^* = (K_\downarrow - K_\uparrow) - (L_\uparrow - L_\downarrow) \tag{14}$$

According to the surface energy balance:

$$Q^* = Q_J + \Delta Q_s = LE_{opt} + H_{opt} + \Delta Q_s = cH_{opt} + \Delta Q_s \tag{15}$$

Further solving equation (12) with equations (M20), (M22), (13) and (15), we get

$$cH_{opt} = R_{l,net} - R_{l,0} = K_d(T_s - T_{dry}) \tag{16}$$

To obtain the turbulent flux $H_{opt}$ using equation (12), we derived the expression $(T_s - T_{dry})/T_s$ in terms of $cH_{opt}$ and input variables using equations (M20), (M22), (M23) and (16) and the expression is given as:

$$\frac{(T_s - T_{dry})}{T_s} = \frac{cH_{opt}}{4L_\uparrow - 3cH_{opt}} \tag{17}$$

We estimate $\Delta Q_a$ in terms of $cH_{opt}$ and input variables from the equations (M16), (15), and (M22)

$$\Delta Q_a = 2cH_{opt} - L_\downarrow \tag{18}$$

Using equations (17), (18) and (10), we get a quadratic equation given as:

$$(1 + 6c)H_{opt}^2 - (8L_\uparrow + 3L_\downarrow)H_{opt} + \frac{4 L_\downarrow L_\uparrow}{c} = 0 \tag{19}$$

Based on the above theory and additional relation given by eq.10, we eliminate the use of surface heat storage to derive the quadratic equation for sensible heat flux (eq. 19)

We solve the equation (19) using the solution of the quadratic equation to get the two values of $H_{opt}$ with solution as

$$H_{opt1}, H_{opt2} = \frac{-B \pm \sqrt{B^2 - 4AC}}{2A} \tag{20}$$

Where:

$A = (1 + 6c)$.

$B = -(8L_\uparrow + 3L_\downarrow)$

$C = \frac{4L_\downarrow L_\uparrow}{c}$

$c = 1 + \frac{s}{\gamma}$

We take the maximum value from the two solutions of the $H_{opt}$ to get the best estimate. The maximum value from the two solutions of $H_{opt}$ is used to maximize the maximum convective power ($G_{d,max}$).

$$H_{opt} = max\{H_{opt1}, H_{opt2}\} \tag{21}$$

We obtain corresponding optimal value of $LE_{opt}$ at maximum convective power from Eq. (7) using the value of $H_{opt}$. Using Eq. (15), we obtain the total turbulent heat flux ($Q_J$).

The abovementioned formulation requires the thermodynamic heating to be sufficient to establish a convective process for maximum power such that $Q^* > R_{lnet}$ i.e $R_s > 2R_{lnet}$. Such conditions cease to exist during, dawn, dusk and night. This may result in the value of $H_{opt}$ greater than $Q^*$, which is impossible for actual conditions during these times. In such cases, we assume $H_{opt} = 0$, due to insufficient availability of net energy. Subsequently, $LE_{opt} = 0$, resulting in $Q^* = \Delta Q_s$ from the SEB.

Thus, the total turbulent energy is given by the sum of $H_{opt}$ and $LE_{opt}$.

Further, we estimate the $T_s$ based on the above results.

Based on the equations 16 and M22, we get:

$$R_{l,out} = L_\uparrow - cH_{opt} \tag{22}$$

From equations 10, 18, 22 and M24, we get:

$$T_s = \left(\frac{cH_{opt} + 4R_{l,out}}{4R_{l,out}}\right) T_{dry} \tag{23}$$

We found the thermodynamically estimated $T_s$ (Now referred as $T_{st}$) from equation 23 is higher than the $T_s$ calculated from $T_s = \left(\frac{L_\uparrow}{\sigma}\right)^{\frac{1}{4}}$. The higher value of $T_{st}$ is due to the conservative estimation of the slope of saturation pressure curve ($s$) (as explained after equation 7) calculated at the temperature $T_r$ ($T_r < T_s$). The use of $T_r$ in the estimation of $s$ results in a lower value of turbulent flux. We compensate for the differential energy ($Q_{Diff}$) in outgoing longwave radiation resulting from the differences between $T_{st}$ and $T_s$. The $Q_{Diff}$ is given as:

$$Q_{Diff} = \sigma T_{st}^4 - \sigma T_s^4 \tag{24}$$

Thus, the total adjusted turbulent flux is now:

$$Q_J = H_{opt} + LE_{opt} + Q_{Diff} \tag{25}$$

And the land surface heat storage flux ($\Delta Q_s$) using equation 25 is given by:

$$\Delta Q_s = Q^* - Q_J \tag{26}$$

We then deduce the sensible and latent heat fluxes at the equilibrium partitioning.

$$H_{opt} = \frac{\gamma}{\gamma + s} Q_J \tag{27}$$

And

$$LE_{opt} = \frac{s}{\gamma + s} Q_J \tag{28}$$

Here, in the equations 27 and 28, the $s$ is calculated at $T_s$ following literature[43,44]. The equation 13 describes the condition in which the evapotranspiration is not limited by the water availability. For actual conditions, a stress factor ($f_w$) is introduced to estimate the actual $LE$ and $H$. These in the water limiting condition are given by the equations [43]:

$$LE = f_w\, LE_{opt} \qquad (29)$$

$$H = Q_J - LE \qquad (30)$$

## Results and Discussions

### Thermodynamic estimate of turbulent flux and its validation

*In-situ point estimates for global sites*

To validate the developed theory, we used the FLUXNET2015[41] database, which comprises in-situ observational data of turbulent flux and radiation components around the globe covering different climate zones and different land covers based on the IGBP classification. We also validated three urban regions limited to data availability (Refer Section B.1 in Methods). We first computed the thermodynamic estimate (TE) of turbulent flux calculated using the expressions (Equations 19-25) (now denoted by $Q_{Jthermo}$) for flux sites, using four radiation components, $K_\downarrow, K_\uparrow, L_\downarrow, L_\uparrow$. The radiation components were taken directly from the FLUXNET sites to estimate $Q_{Jther}$. We then compared $Q_{Jthermo}$ with the observed turbulent flux by eddy covariance (EC) technique ($Q_{JEC}$), at 102 sites (99 from FLUXNET and 3 urban regions) for which data of all radiation components and $Q_{JEC}$ were available. We present the results for 38 sites, which have a minimum of 30 months of data (Extended Figure 1, shown in map). For the urban regions, there are only 3 sites available, and we considered all of them. In extended Figure 2, we have shown diurnal variation in $T_{st}$ and $T_s$ that leads to differential energy which is corrected in the total turbulent convective energy (Eq. 24). In Figure 2, we present the average monthly diurnal variations of turbulent flux, comparing $Q_{Jthermo}$ and $Q_{JEC}$.

Figure 2 shows that turbulent heat flux thermodynamic estimates follow the diurnal variations depicted by the eddy covariance observations. The errors are minimal for the forest regions, Deciduous Broadleaf Forest (DBF), Evergreen Broadleaf Forest (EBF) and Evergreen Needleleaf Forest (ENF) [first 3 rows of Figure 2]. For the wetland (WET), at the site, US-WPT, the $Q_{JEC}$ value is quite low compared to $Q_{Jthermo}$. Such a low value of $Q_{JEC}$ could be associated with measurement limitations, as the turbulent heat fluxes are normally high for wetlands due to high latent heat flux. Good resemblance between $Q_{Jthermo}$ and $Q_{JEC}$ are observed in other WET sites in Figure 2. The plots for $Q_{Jthermo}$ and $Q_{JEC}$ are very similar for cropland (CRO) and Savannas (SAV) (Figure 2). There are discrepancies for a few sites for grassland (GRA) and Open shrublands (OSH). The eddy-covariance observations at those sites show high closure terms; hence, there may be a possibility of measurement limitations. For urban regions (URB), there are differences, which may be because of urban structure that often introduces errors in $Q_{JEC}$. Overall there is a very good match between $Q_{Jthermo}$ and $Q_{JEC}$, showing the efficacy of the thermodynamic model. The summary of RMSE (Root Mean Square Error) and MBE (Mean Bias error) of 102 sites is presented in Supplementary Table1.

Extended Figure 3 presents the scatter plots of monthly values between $Q_{Jther}$ and $Q_{JEC}$ at individual sites. For the majority of sites for land use, DBF, EBF, ENF, WET and CRO, the points closely fall on the $45^0$ lines with a very high $R^2$ value between $Q_{Jthermo}$ and $Q_{JEC}$. There are deviations for GRA,

SAV, OSH and Urban sites, which are consistent with our observations from Figure 2. We regressed $Q_{Jthermo}$ against $Q_{JEC}$ and presented the slope (*m*) and intercept (*c*) in Extended Figure 2. For most cases, the slope is close to 1, and the intercept value is low, showing similarities between the thermodynamic estimates and on-site observations. Supplementary Table 1 summarizes the analysis of regression performances and coefficients (mean ± Standard Deviation (SD)) for all 102 sites. The adjusted $R^2$ value of 0.86 (Extended Table1) shows good agreement between $Q_{Jther}$ and $Q_{JEC}$. The thermodynamic approach developed here is able to adequately explain the monthly variability of on-site turbulent heat flux observations. 56 out of 102 sites show $R^2$ values greater than 0.9, and 82 sites greater than 0.8. The slope and intercept describe the underestimations and overestimations by $Q_{Jthermo}$ with respect to $Q_{JEC}$. Overall, the analysis reveals slightly greater estimates of $Q_{Jthermo}$ for larger values of $Q_{JEC}$ based on an average slope of 1.12. Further, a low mean average intercept of 1.9 W m$^{-2}$ indicates low bias for lower values. The discrepancies between the estimates, $Q_{Jthermo}$ and $Q_{JEC}$ indicate either the limitation of TE estimations or EC estimations, or both. The detailed literature review[20] of the past 25 years discussed the limitation of EC observations. They reported systematic underestimations of the turbulent energy due to uncertainties in the complicated data analysis in the EC technique. Further, the observations by a single flux tower are unable to capture larger-scale mesoscale eddy circulations. At almost all the flux tower sites, the surface energy balance does not follow the conservation of energy since the available energy ($Q^*$) is more than the sum of observed variables, $Q_{JEC}$ and $\Delta Q_s$ measured by heat flux plate, resulting in residual energy ($Q_{Res}$). It could be possible that the TE estimates are overcoming the limitations of EC techniques; however it is difficult to infer the same from the present analysis.

A detailed study[19] based on the analysis of 173 FLUXNET sites concluded that the differences in the surface energy balance closure ($C_{EB}$), reciprocal to $Q_{Res}$, between different land covers (forests, non-forests, and other areas) are insignificant. However, the literature suggests that the improvement in the $C_{EB}$ is possible addressing underestimation of soil heat storage[45] and considering unquantifiable factors, such as the role of water[19], heat storage in the canopy[31] and metabolic terms, as well as photosynthesis[46]. Further, a study[47] which assesses the energy balance closure showed that the heat flux from tree biomass, ignored in energy balance, is the biggest of the anticipated storage components in $\Delta Q_s$. The case study[31] of modelling turbulent fluxes for shallow vegetation showed improvement by including canopy heat storage elements. Thus, the $Q_{Res}$ from EC estimation is also associated with the underestimation of $\Delta Q_s$ and should be dependent on the land cover types for obvious reasons. Further, as the $Q_{Res}$ ($C_{EB}$ is lower) is higher during daytime with higher $Q^*$ [19,48], Figure 2 shows higher biases between the $Q_{Jthermo}$ and $Q_{JEC}$, during daytime, which could be due to high $Q_{Res}$.

To understand the dependence of the bias on land cover, we present the relative differences in turbulent heat fluxes between $Q_{Jthermo}$ and $Q_{JEC}$, with respect to $Q^*$ (scaled by $Q^*$ for comparison) in Extended Table 2. The biases are computed for the peak daytime (10:00-14:30 local standard time). We applied the two-sample Kolmogorov-Smirnov two-sided test and k-sample Anderson-Darling test at a significance level of 0.05 to understand the differences between the bias samples from different land covers. We found that the scaled bias for forests (2.6±13.5%) is significantly different from non-forests (includes GRA and CRO, 10.1±8.7%), Wetlands (40.6±64.2%), Urban areas (30.5±1%), and Other areas (includes SAV and OSH, 13.3±8.9%). Further, the individual differences are significant between non-forest & wetlands (WET), and Others and WET. Based on the monthly analysis, we found that the biases for the sites in non-forests (CRO and GRA) and Others (SAV and OSH) are highest in the precipitation months (Extended Figure. 4). This bias is probably due to the limitations of EC stations in estimating high Latent Energy (LE). Thus, we conclude that $Q_{Jthermo}$ captures the limitation of $Q_{JEC}$ to observe high turbulent energy, which occurs mainly in canopies with low biomass and heat storage elements. We found higher biases or additional turbulent flux compared to EC measurements for areas

*Global estimates of turbulent heat fluxes and their validation*

After validating our methodology, we used it to develop globally distributed land surface estimates of turbulent fluxes at 1° spatial resolution using four radiation variables, all-sky incoming and outgoing shortwave and longwave radiation flux ($K_\downarrow, K_\uparrow, L_\downarrow, L_\uparrow$), at the surface level obtained from the remotely sensed dataset CERES Edition4A SYN1deg-MHour, which is at a spatial resolution of 1° and temporal resolution of monthly hourly (Refer to Section B.2 in Methods). We present the results in Figure 3. The figure shows the spatial variations of the radiation components at the surface for four seasons, DJF, MAM, JJA, and SON. The time period considered is 2003-2019. The globally distributed land surface Net radiation ($Q^*$) (Figure 3, top row) estimated from the incoming and outgoing fluxes, is used to calculate the global turbulent heat flux field ($Q_J$)(Figure 3, 2$^{nd}$ row) based on thermodynamic principles using the expressions (19-25). Figure 3, 3$^{rd}$ row presents the land surface storage heat flux($\Delta Q_S$). The major parts of the northern hemisphere have the highest positive value of ($\Delta Q_S$) in JJA, followed by MAM, which steadily becomes negative in SON and highest negative during DJF. This is expected due to seasonal patterns. However, we found positive $\Delta Q_S$ throughout all the seasons for Amazon forests and Middle Eastern Africa characterized by mostly Tropical and Subtropical moist climates with moist broadleaf forests, grasslands, savannas and shrublands. We further calculated the global latent and sensible heat flux fields (Figure 3, last two rows) using equations 27-30. The evaporative stress factor ($f_w$) is obtained from the GLEAM v3.6b dataset[49,50].The $f_w$ is based on vegetation optical depth and root zone soil moisture obtained using remote sensing and used in land evaporation products.. During the JJA season, the monsoon regions north of the Equator, such as South Asia, have a high latent heat flux due to the wet season. The desert/ arid regions, like Sahara in Africa or California in the US, have low latent heat flux, resulting in very high sensible heat flux. The radiation components are in general lowest in the northern hemisphere in DJF due to seasonally low $Q^*$.

We validated our estimates using remote sensing data with the flux tower estimates. We picked up the site with the highest data points for validation for each land cover. We see the monthly variation between CERES and FLUXNET by assessing variation for absolute magnitude (Extended Figure 5) and monthly anomalies after deseasonalisation (Extended Figure 6). For absolute values, we have found a very strong similarity. The results depict high agreement in the variables, $K_\downarrow, L_\downarrow$ and $L_\uparrow$ of CERES and FLUXNET 2015 in-situ dataset with adjusted R$^2$ values greater than 0.96 for all the sites. However, we observe an underestimation of $K_\uparrow$ from CERES for a few sites. The $Q_{Jthermo}$ estimated from these input variables from CERES shows high agreement with an adjusted R$^2$ greater than 92%. The scatter plots between thermodynamic estimates from satellites and in-situ observations show all the points falling close to the 45$^0$ line. For monthly anomalies, the CERES shows good agreement for $K_\downarrow, L_\downarrow$ and $L_\uparrow$ to FLUXNET but less as compared to the absolute values. The $K_\uparrow$ shows less agreement as it depends on land surface characteristics like albedo and land use, which differs due to scaling challenges while linking remote sensing footprints to tower footprints. Further, the estimated $Q_{Jthermo}$ from CERES matches well (with average adjusted R$^2$ = 0.6) with the FLUXNET sites observations. Thus, we see the capability of the analytical thermodynamic-based expressions to estimate the global surface fluxes.

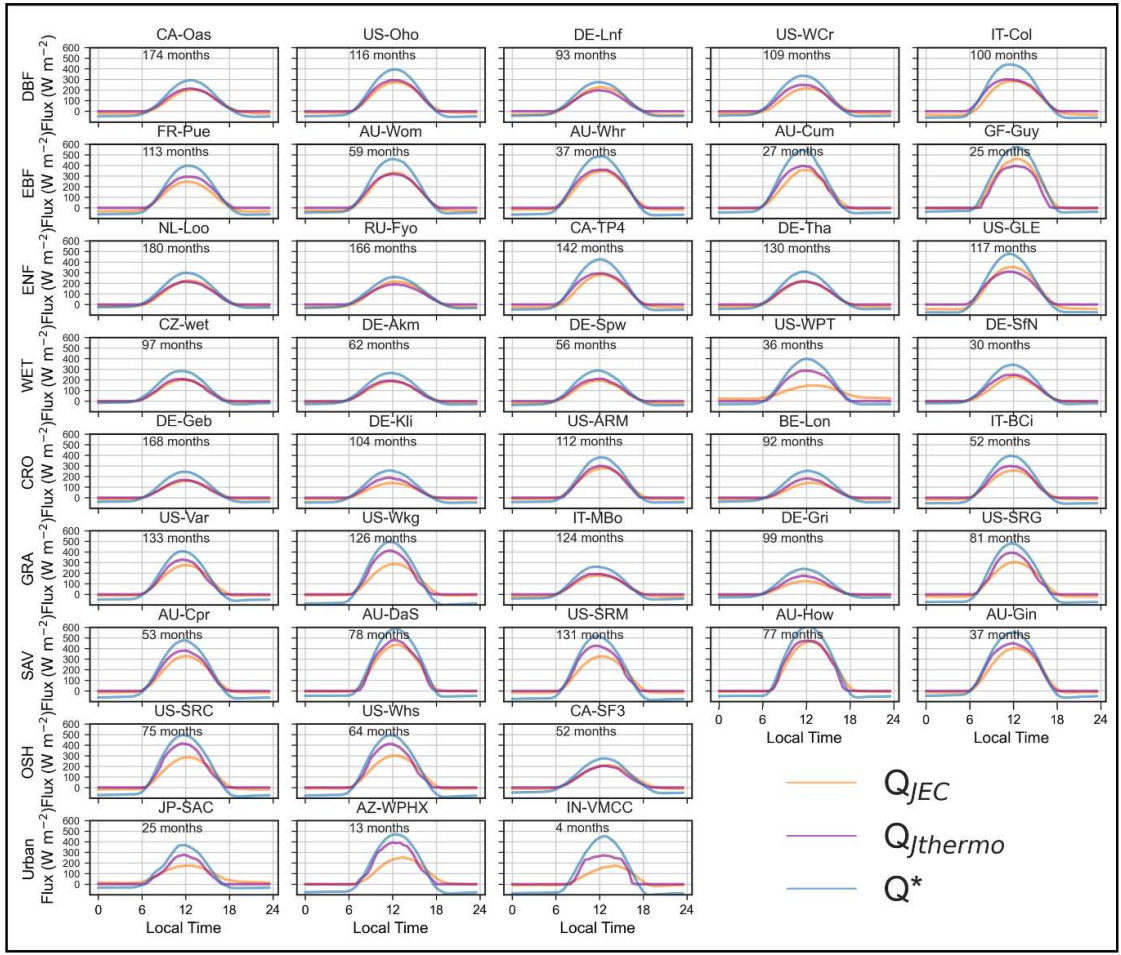

Figure 2. Comparing the average diurnal variations of the fluxes from TE theory and EC observations for 38 FLUXNET sites and three urban sites. They are arranged based on land covers: DBF Deciduous Broadleaf Forest, EBF- Evergreen Broadleaf Forest, ENF- Evergreen Needleleaf Forest, WET- Wetlands, CRO- Croplands, GRA- Grasslands, SAV- Savannas, OSH- Open Shrublands, and Urban. The locations of the sites are presented in Extended Figure 1. $Q_{JEC}$ - Observed turbulent heat flux by EC, $Q_{Jthermo}$ - turbulent heat flux estimated from the thermodynamic model, $Q^*$ - Observed net radiation.

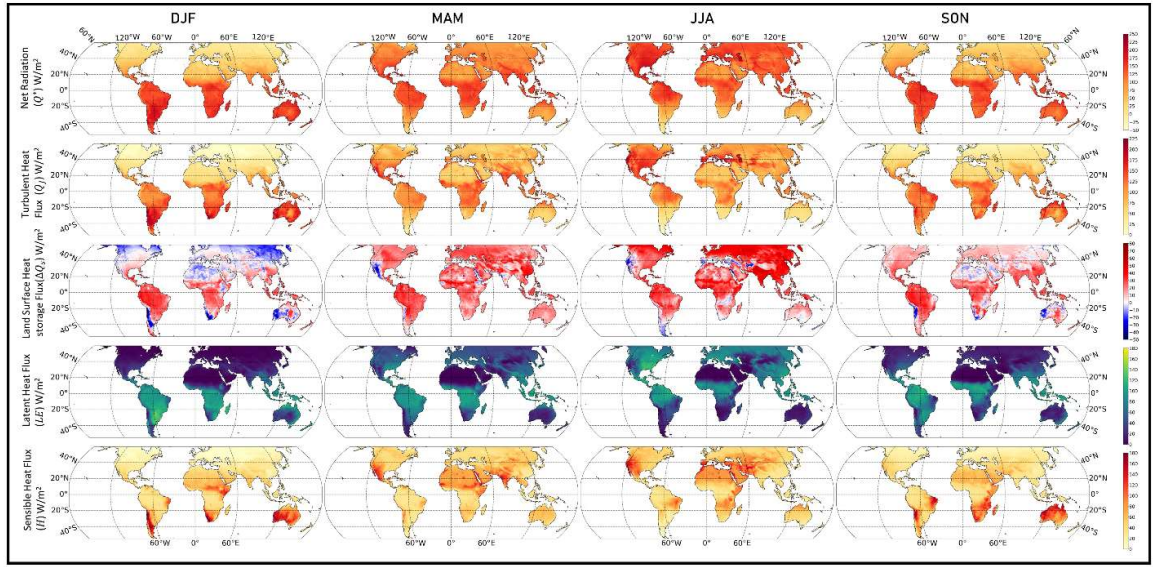

*Figure 3: The annual average fluxes for 2003-2019 are estimated from the CERES monthly global Syn dataset at 1˚ resolution. $Q^*$ (Net Radiation) derived from CERES radiation components ($K_↓$, $K_↑$, $L_↓$, $L_↑$). $Q_J$ (Turbulent heat flux) and $\Delta Q_S$ (land surface heat storage) are the estimated fluxes from the thermodynamic model. The global latent heat (LE) and Sensible heat (H) flux fields are estimated based on the evaporative stress factor from the GLEAM dataset.*

*Thermodynamic estimate of global land surface heat storage*

The land surface heat storage flux ($\Delta Q_S$) is defined as the net heat storage change (uptake or release) in the volume per unit horizontal land area. The land volume comprises a certain depth of the ground and the canopy elements associated with land covers such as buildings and vegetation in urban areas and biomass of trees in forests[32]. Here, we present the first global estimates of $\Delta Q_S$ as energy balance residual using equation 1 with $Q_J = Q_{Jther}$ and analyze its behaviour geographically. In general, $\Delta Q_S$ is minimal in wet areas with sparse canopy elements due to high LE but reaches a higher value in arid regions with values equivalent to $H$ and often exceeds $LE$. Further, $\Delta Q_S$ with a considerable amount of biomass heat storage in forest regions dampens the diurnal temperature range[28]. We examine these behaviours of $\Delta Q_S$ on global land by estimating daily means (Figure 4, column 1) and daytime means (10:00 to 14:30 local time, Figure 4, column 2) for 2003-2019 to see if these processes are accurately represented. The arid regions such as mainly in the southwestern United States, the Sahara desert in northern Africa, the Arabian Peninsula, the Thar desert in India, and central Australia have a very high positive $\Delta Q_S$ (Figure 4c) and $\Delta Q_S/Q^*$ (Figure 4d) during the peak solar hours because of low LE. However, in the non-peak solar hours the land surface in these regions releases much heat, resulting in very low to negative 24hrs mean values of $\Delta Q_S$ (Figure 4a) and $\Delta Q_S/Q^*$ (Figure 4b). The mean daily $\Delta Q_S$ and $\Delta Q_S/Q^*$ show a substantial amount of energy throughout the Amazon forests compared to barren regions like the Sahara desert. This characteristics is due to the heat storage capacity of the biomass. However, during the peak solar hours, high evapotranspiration and high LE results in a comparatively low land heat storage flux, as compared to the arid regions. Further, we see that the percentage $\Delta Q_S/Q^*$ is highest in the arctic-boreal regions (Figure 4(c)) during both daily and daytime mean as permafost acts as a large sink of energy and net radiation melts ice in the active layer[51]. However, $\Delta Q_S$ during peak solar hours shows smaller values than arid regions due to the low net radiation (Figure 4(d)).

With an absorbed land-mean solar radiation of 143 Wm$^{-2}$ at the surface and net longwave emitted from the land surface given by 59 Wm$^{-2}$, we estimate the global land (90°N-90°S) $Q^*$ of 84 Wm$^{-2}$ for 2003-2019 (Extended Table 3) based on the CERES Syn dataset. The estimate is similar to the CERES EBAF $Q^*$ of 79.10 Wm$^{-2}$ calculated for the same period. Our estimates are slightly higher than the estimated $Q^*$ values of 77.5 Wm$^{-2}$ by Jung et al.[18], 76 Wm$^{-2}$ by L'Ecuyer et al.[52], and 70 Wm$^{-2}$ Wild et al.[53]. Based on the $Q^*$ as input from CERES Syn, the thermodynamic model estimates global land $Q_J$ as 82 Wm$^{-2}$ and $\Delta Q_s$ as 2 Wm$^{-2}$. All the global value in this study are determined as area-weighted average over the land. The global LE is 40 Wm$^{-2}$. The value is consistent and in good agreements with 39.5 W m$^{-2}$ by Jung et al.[18], 38.5 W m$^{-2}$ by Trenberth et al.[54], 37-59 W m$^{-2}$ by Jiménez et al.[55], and 38 W m$^{-2}$ by Wild et al.[53]. For the same global spatial extent global H, as the difference between $Q_J$ and LE, is 42 W m$^{-2}$. The estimated H is in excellent agreement with Jung et al.[56] with the value of 41± 4 Wm$^{-2}$, higher than the range of 36–40 W m$^{-2}$ estimated by Siemann et al.[57], lies well above 27 W m$^{-2}$ estimated by Trenberth et al.[54], and well within the range of 18-57 W m$^{-2}$ estimated by Jiménez et al.[55].

We found that the land use land cover drives the distribution of land surface heat fluxes (Extended Table 3). For example, the forest regions have a high value of global area averaged $\Delta Q_s$ for the period 2003-2019 as 6.72 Wm$^{-2}$. This is because of high canopy heat storage potential. On the contrary, the non-forest regions have a globally averaged $\Delta Q_s$ as 2.7 Wm$^{-2}$. The same is true for shrublands and Savannas with average value of 1.5 Wm$^{-2}$. The barren regions have negative values of -2.47 for $\Delta Q_s$. The distributions show the role of vegetation in storing heat fluxes and thus driving the partitioning of incoming radiation fluxes perturbing local climate. The differential values of $\Delta Q_s$ across land use land covers are consistent with previous literature on parametrized models for heat storage[28].

## **Conclusion**

We developed an analytical approach to estimate turbulent and land surface heat storage fluxes based on the thermodynamic principles of maximum convective power. Working on the previous studies on the thermodynamic theory that describes the land-atmosphere as a radiative-convective system, we improved the theory to estimate the turbulent fluxes successfully. The uniqueness of the approach is that this needs only incoming and outgoing surface radiation fluxes that are merged satellite and model products. The methodology is validated against the flux tower observations across the globe from different land use land covers. For the first time, such a method provides globally gridded analytical estimates of turbulent and land surface storage heat fluxes without using climate or land surface models involving parameterizations, thus not suffering from model deficiencies and uncertainties. We found that such thermodynamic estimates also overcome some of the limitations of the eddy covariance estimates. We further analyzed the spatial and diurnal variations of the turbulent and land surface heat storage fluxes and found them consistent with our existing site-specific knowledge. The methodology is applicable to in-situ observation sites as well as to a large region. The surface energy products generated through the present study overcome the limitations of the non-existence of observed surface energy flux data and will help the climate community understand the trajectory of surface processes across land use changes in a warming environment.

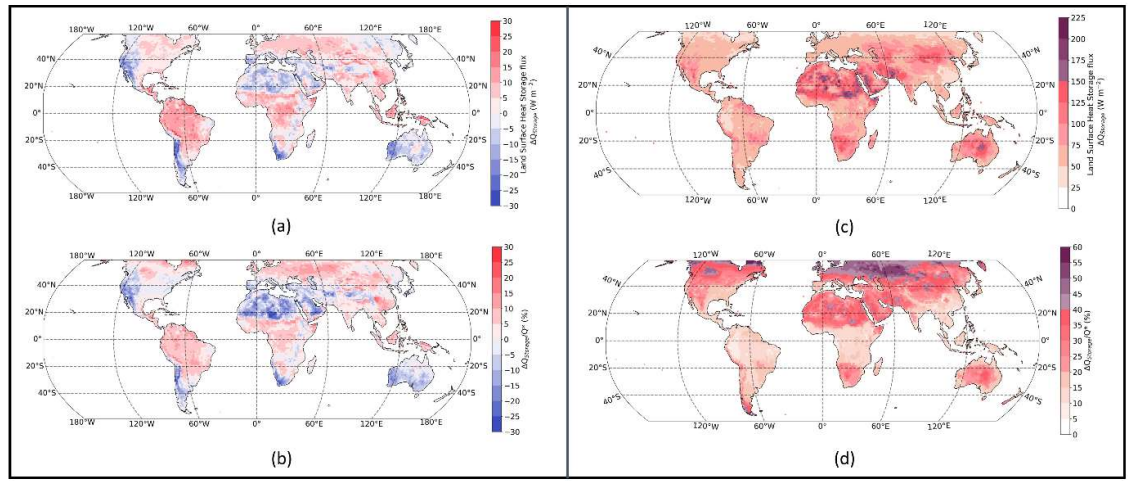

*Figure 4: Global land surface heat storage fluxes estimated using the thermodynamic model derived in this study. Column 1 depicts the daily means of (a) Land surface heat storage flux $\Delta Q_S$ and (b) $\Delta Q_S$ scaled by Q\* ($\Delta Q_S/Q^*$). Column 2 depicts the same variables (c) $\Delta Q_S$ and (d) $\Delta Q_S/Q^*$ but averaged over local daytime (10:00-14:30 HRS).*

# Methods:

## (A) Atmosphere as a convective dissipative heat engine

The radiative heating of the surface makes the air parcel in contact gain a lower density, increase its potential energy, and develop a state of thermodynamic disequilibrium. This disequilibrium causes the system to derive work as a heat engine to generate a convective motion by creating a buoyant force to reduce the potential energy of the parcel and depletes the temperature gradient. This motion is associated with transporting energy, mass, and momentum through fluxes between the surface and the atmosphere.

**(A.1) Carnot limit of a heat engine:** Consider a heat engine with two reservoirs, a hot reservoir and a cold reservoir, and heat fluxes from the hot reservoir as $J_{in}$ and comes out from the cold reservoir as $J_{out}$. In the case of the land-surface atmospheric convective system similar heat engine develops such that the hot reservoir is the surface, a source of turbulent fluxes $J_{in}$, with a temperature $T_s$. The cold reservoir is at the boundary of the atmosphere, where the effect of local convection merges with large-scale motion, and the engine releases heat $J_{out}$ from the local convective system at a radiative temperature $T_r$. The engine develops a power $G = dW/dt$, work per unit time, that generates convective motion using turbulent heat as input from the surface. For maximum Carnot power and efficiency, no heat is used to increase the engine's internal energy, and the heat input is fully utilized to generate power.

The energy balance is expressed as:

$$J_{in} = J_{out} + G \tag{M1}$$

The entropy balance of the engine with internal entropy generation ($\sigma$) is:

$$\sigma + \left(\frac{J_{in}}{T_s} - \frac{J_{out}}{T_r}\right) = 0 \tag{M2}$$

To derive power $G$, the expression of $J_{out}$ from the energy balance (eq.M1) is inserted into the entropy balance (eq.M2).

$$G = J_{in}\left(\frac{T_s - T_r}{T_s}\right) - \sigma T_s \tag{M3}$$

According to the second law of thermodynamics, $\sigma \geq 0$. The maximum power ($G_{carnot}$) is achieved when there is no entropy generation inside the engine $\sigma = 0$. So, the expression of $G_{carnot}$ is

$$G_{carnot} = J_{in}\left(\frac{T_s - T_r}{T_s}\right) \tag{M4}$$

The Carnot efficiency, maximum efficiency due to the maximum power, is given by:

$$\eta_{carnot} = \frac{G_{carnot}}{J_{in}} = \left(\frac{T_s - T_r}{T_s}\right) \tag{M5}$$

**(A.2) Power of a dissipative heat engine:** In a dissipative engine, the generated mechanical work dissipates within the engine and increases its internal energy. The dissipative heat in such an engine, even if it dissipates near the hot reservoir, does not act as an additional source with input heat, $J_{in}$. It increases the internal energy of the system. The radiative-convective process of the land atmosphere behaves as a dissipative engine. Considering the same engine as above but with an additional process of frictional dissipation ($D$) and additional term, we get the state of internal energy of the system ($\Delta U$). The power generated for the convection motion in the engine is given by $G$.

According to the first law of thermodynamics, the energy budget of a heat engine:

$$\Delta U = \Delta J - G \tag{M6}$$

Where, $\Delta U$ is the internal energy of the engine, $\Delta J$ is the amount of heat flux, and $G$ is the power develops for mechanical work.

For a dissipative heat engine in a steady state, $D$ is added because the total power for convective motion is dissipated as heat within the engine.

$$\Delta U = J_{in} - J_{out} - G + D \qquad (M7)$$

In a steady state, $G = D$ because the total power for convective motion is dissipated as heat within the engine. In addition, $G = D = \Delta U$ because the generated heat from frictional dissipation cannot be used as an additional heat source to generate work but converts into the internal energy of the engine to raise its temperature. So, equation (M7) becomes:

$$\Delta U = J_{in} - J_{out} = D = G \qquad (M8)$$

The associated entropy budget of a dissipative heat engine is given by the entropy associated with the change in the internal energy of the engine at an effective temperature of the heat engine $T_a$, entropy associated with the input energy from the hot reservoir at the temperature $T_s$, entropy from the emitted heat at the temperature $T_r$, entropy due to frictional dissipation term $D$ at $T_a$, and the irreversible entropy production ($\sigma_{irr}$) within the engine other than due to frictional dissipation.

$$\frac{\Delta U}{T_a} = \frac{J_{in}}{T_s} - \frac{J_{out}}{T_r} + \frac{D}{T_a} + \sigma_{irr} \qquad (M9)$$

For a system where the irreversible entropy generates only due to the frictional dissipation, entropy through non-frictional dissipation is zero ($\sigma_{irr} = 0$). In that case, the maximum power or Carnot limit of the dissipative heat engine is estimated by assuming $\sigma_{irr} = 0$ in the (M9), eliminating $J_{out}$ in equation (M9) with the expression from equation (M8), and using $G = D$ in steady-state.

$$G = J_{in} \frac{T_a}{T_s} \cdot \frac{T_s - T_r}{T_r} - \Delta U \cdot \frac{T_a - T_r}{T_r} \qquad (M10)$$

For the land-atmosphere dry convective system, $T_a$ is the mean temperature of the atmosphere. As the convective engine is operated in the lower atmosphere, it is closer to the surface temperature and can be reasonably assumed as $T_a \approx T_s$. The atmospheric heat storage represents the internal energy of the engine ($\Delta U = \Delta Q_a$). For a dry convective system, the Sensible heat represents the heat source ($J_{in} = H$), and the temperature of the cold reservoir is given by the dry sink temperature $T_{dry}$. Thus, from (M10), we get the expression of the limit on power generation of the dissipative convective engine in which irreversible entropy generates only because of the irreversible frictional dissipation.

$$G_d = (H - \Delta Q_a) \cdot \frac{T_s - T_{dry}}{T_d} \qquad (M11)$$

Further, from equation (M7), $G = D = \Delta Q_a$, equation (M11) becomes:

$$G_d = H \cdot \frac{T_s - T_{dry}}{T_s} = \Delta Q_a \qquad (M12)$$

**(A.3) The storage heat distribution in the land-atmosphere dry convective system**

The energy budget equation of the surface is given by:

$$\Delta Q_s = R_s - R_{l,net} - H - LE \tag{M13}$$

The energy budget of the atmosphere is given by:

$$\Delta Q_a = R_{l,net} + H + LE - R_{l,out} \tag{M14}$$

And the entropy budget is given by:

$$\frac{\Delta Q_a}{T_e} = \frac{H}{T_s} + \frac{LE}{T_{dry}} + \frac{R_{l,net}}{T_{dry}} - \frac{R_{lout}}{T_{dry}} + \frac{D_k}{T_e} + \Delta S_{dif} \tag{M15}$$

$T_e$ is the temperature of the engine. The $T_e$ is approximated by $T_s$ as the process is taking place close to the surface. The total energy balance of the system and heat storage ($\Delta Q_{total}$) in the land-atmospheric system during the dry convection is given by adding equations (M13) and (M14):

$$\Delta Q_{total} = \Delta Q_a + \Delta Q_s = R_s - R_{lout} \tag{M16}$$

According to equation (M12), we assumed that the heat storage in the atmosphere ($\Delta Q_a$) or an increase in the internal energy of a dry convective system only occurs due to the changes within the heat engine. However, a parallel radiative transfer of energy $R_{l,net}$, independent of the heat engine, also contributes to heat storage changes in the atmosphere. It is essential to understand the change in the power limit given by equation (12) due to $R_{l,net}$. According to the study[40], in both cases where the atmosphere is opaque to $R_{l,net}$ or completely transparent does not change the power limit given by the equation (M12).

### (A.4) Linearization of longwave radiative heat transfer ($R_{l,net}$)

$$R_{l,net} = L_\uparrow - L_\downarrow \tag{M17}$$

$$R_{l,net} = \sigma T_s^4 - \frac{3}{4}\tau\sigma T_{dry}^4 \tag{M18}$$

The surface is assumed as a blackbody with thermal emission given by $\sigma T_s^4$. $\sigma$ is the Stefan-Boltzmann constant ($\sigma = 5.67 \times 10^{-8} \text{Wm}^{-2} K^{-4}$). $L_\downarrow = \frac{3}{4}\tau\sigma T_{dry}^4$ gives the longwave radiative flux towards the surface. $\tau$ is the longwave optical depth (or thickness) of the atmosphere. In our model, we use $L_\downarrow$ as the input. However, many parameterization schemes exist to estimate $L_\downarrow$ from on-ground meteorological variables[58,59]

To estimate maximum convective power, the expression of net longwave radiation is linearized using first-order Taylor expansion around the sink temperature ($T_{dry}$).

$$\sigma T_s^4 \approx \sigma T_{dry}^4 + K_d(T_s - T_{dry}) \tag{M19}$$

Where $K_d$ is a first order constant given by $K_d = 4\sigma T_{dry}^3$

Combining equations (M17), (M18), and (M19), we get the linear approximation of $R_{l,net}$ in terms of temperature difference.

$$R_{l,net} = R_{l,0} + K_d(T_s - T_{dry}) \tag{M20}$$

With constant $R_{l,0}$ given as:

$$R_{l,0} = \sigma T_{dry}^4 - L_\downarrow \quad (M21)$$

The atmospheric temperature at which the heat is radiated out from the dry convective engine is the effective dry sink temperature ($T_{dry}$). It is the temperature with the highest radiative entropy to radiate out from the dry convective heat engine with associated emission of radiation given by radiative flux ($R_{l,out} = \sigma T_{dry}^4$).

So equation (M21) becomes:

$$R_{l,0} = R_{l,out} - L_\downarrow \quad (M22)$$

$$K_d = \frac{4 R_{l,out}}{T_{dry}} \quad (M23)$$

And,

$$T_{dry} = \left(\frac{R_{l,out}}{\sigma}\right)^{\frac{1}{4}} \quad (M24)$$

**(A.5) Turbulent fluxes from the maximum convective power limit:** The dry convection in the lower atmosphere transports heat as Sensible heat flux and passively transports latent heat flux until water vapour condenses. The surface also cools through the radiative transfer of heat $R_{l,net}$. In a complete land-atmospheric convective system boundary, the heat engine and $R_{l,net}$ reduces the temperature difference of boundaries. The sensible heat flux in dry convection is derived by maximizing the convective power $\left(\frac{dG}{dH} = 0\right)$. The equation (M13) from equation (M20) can be written as:

$$\Delta Q_s = R_s - R_{l,0} - K_d(T_s - T_{dry}) - H - LE \quad (M25)$$

From equation (M25), we can express the temperature difference as:

$$T_s - T_{dry} = \frac{R_s - R_{l,0} - H - LE - \Delta Q_s}{K_d} \quad (M26)$$

From equation 7 in theory section, we can replace $LE_{opt}$ in terms of $H_{opt}$ in equation (M26)

$$T_s - T_{dry} = \frac{R_s - R_{l,0} - H\left(1 + \frac{S}{\gamma}\right) - \Delta Q_s}{K_d} \quad (M27)$$

To obtain the maximum convective power of the engine driven by the sensible heat flux ($H$), $G_d$ from the equation (12) and (M27) can be written as:

$$G_d = H \cdot \frac{R_s - R_{l,0} - H\left(1 + \frac{S}{\gamma}\right) - \Delta Q_s}{T_s K_d} \quad (M28)$$

**(B) Data and processing**

**(B.1) Eddy Covariance observations**

We used the observations of turbulent and radiation fluxes from the FLUXNET2015 database[41], which comprises of the eddy covariance observations of global sites, non-urban regions, at 0.5h resolution. We used 99 sites based on the availability of all radiation variations, incoming and outgoing longwave and shortwave flux. For total turbulent flux ($Q_{JEC} = H + LE$), we used the sensible and the latent heat flux not corrected for the surface energy balance closure. To ensure the data quality, we used the datapoints (0.5h data) only for which values of all the variables existed. We used total of 6617 site-months of all sites for the validation with minimum of 12 months and maximum of 180 months of a site from FLUXNET data. For urban regions, we used only 3 sites, JP-SAC in Sakai, Japan(25 months)[60], AZ-WPHX in Phoenix, Arizona(13 months)[61], and IN-VMCC in Mumbai, India(4 months)[62] based on the availability of data with us.

**(B.2) Global Satellite data**

We used four radiation variables (incoming shortwave, reflected shortwave, incoming longwave and outgoing longwave radiation) at the surface level as global satellite inputs from the CERES Edition4A SYN1deg-MHour product[63] dataset ( ). The dataset is taken for the study period 2003-2019 and is at a spatial resolution of 1° × 1° degree and a temporal resolution of monthly-hourly. Further, we used the adjusted and all-sky conditions dataset. The surface radiation variables in SYN1deg products are computed based on Langley Fu-Liou radiative transfer model with inputs from the MODIS and CERES geostationary satellites (GEO), GEOS atmosphere and skin temperature, MATCH aerosol constituents, and MODIS spectral aerosol optical depths. The satellite instruments are calibrated against the MODIS data. The adjusted variables are the values constrained to the observed CERES TOA fluxes. In improved Ed4 products, the CERES cloud algorithm provides the cloud properties four times a day to generate hourly resolution. Generally, the Root mean square (RMS) difference of monthly mean Ed4.0 SYN1deg fluxes are similar to the RMS difference of monthly mean Ed4.0 EBAF-Surface fluxes.

We also used the MODIS MCD12C1.006[64] dataset for IGBP global land covers to assess the global surface energy fluxes for land covers. The Land cover data from MODIS was rescaled to the resolution of CERES for the estimation and the analysis.

**Data Availability**

The FLUXNET2015 data used in the study are available at https://fluxnet.org/data/fluxnet2015-dataset/ (https://doi.org/10.1038/s41597-020-0534-3). The CERES Edition4A SYN1deg-MHour product is available from https://ceres.larc.nasa.gov/data/. MODIS MCD12C1.006[64] dataset for IGBP global land covers can be accessed at https://lpdaac.usgs.gov/products/mcd12c1v006/ (https://doi.org/10.5067/MODIS/MCD12C1.006). The GLEAM v3.6b dataset can be accessed from the https://www.gleam.eu/.

**Acknowledgements**

The authors would like to thank the Ministry of Human Resource Development (MHRD), Government of India for funding this study under the grant titled Frontier Areas of Science and Technology (FAST), Centre of Excellence in Urban Science and Engineering (Grant Number 14MHRD005). The work is financially supported by the Department of Science and Technology Swarnajayanti Fellowship Scheme, through project no. DST/ SJF/ E&ASA-01/2018-19; SB/SJF/2019-20/11, and Strategic Programs, Large Initiatives and Coordinated Action Enabler (SPLICE) and Climate Change Program through project no. DST/CCP/CoE/140/2018. SG acknowledges Sonia Seneviratne of ETH, Zurich, and Vishal Dixit o IIT Bombay for technical discussions.

**Extended figure:**

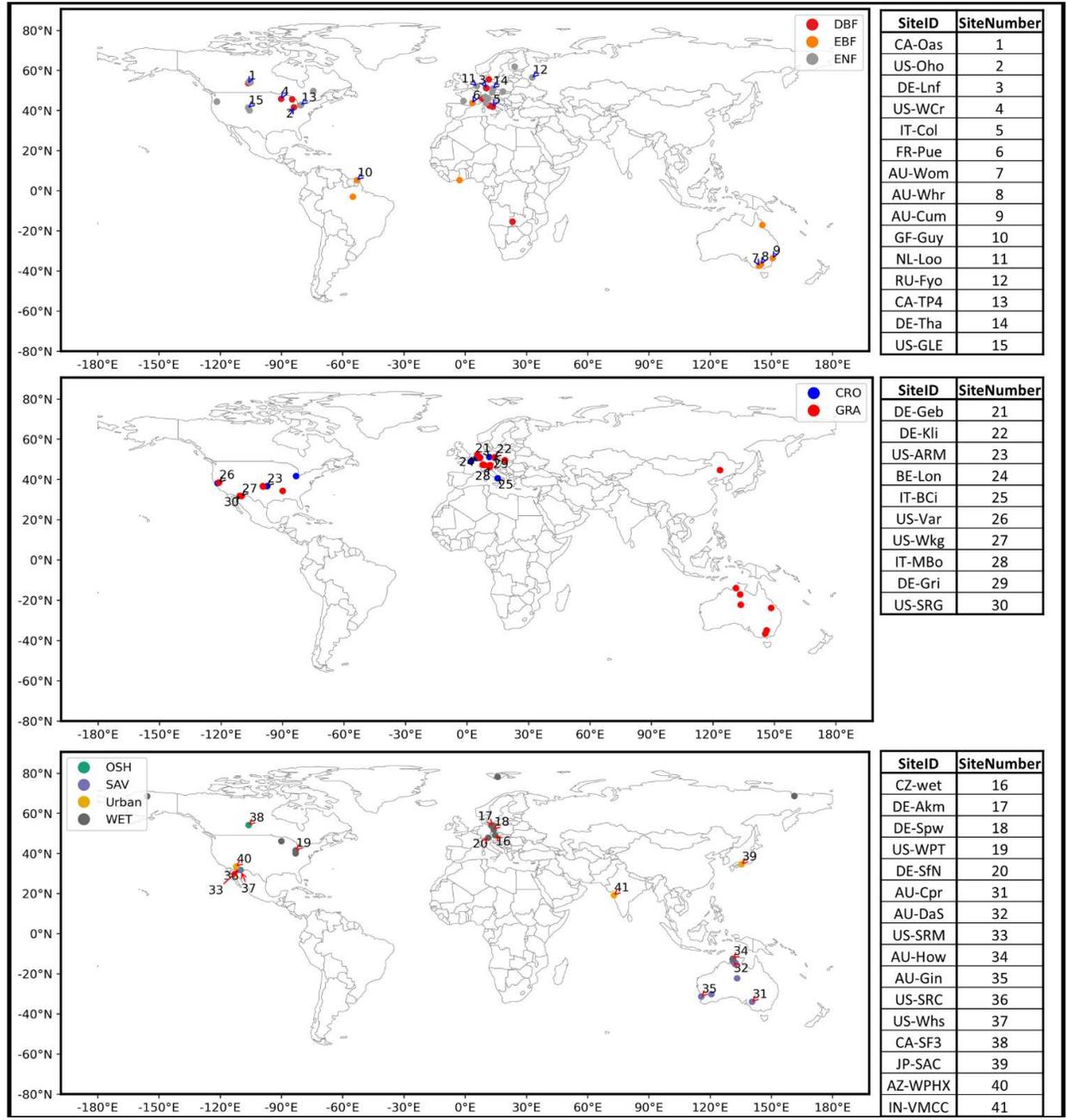

*Extended Figure 1: The geographical locations of all eddy covariance sites used in this study based on IGBP land cover classification of FLUXNET sites (99 sites) and urban stations (3). The tables show only the names with site numbers marked in the maps of the sites used in Figure 2. The figure considers 9 LULC classifications. They are: DBF - Deciduous Broadleaf Forest, EBF- Evergreen Broadleaf Forest, ENF- Evergreen Needleleaf Forest, CRO- Croplands, GRA- Grasslands, SAV- Savannas, OSH- Open Shrublands, Urban, and WET- Wetlands.*

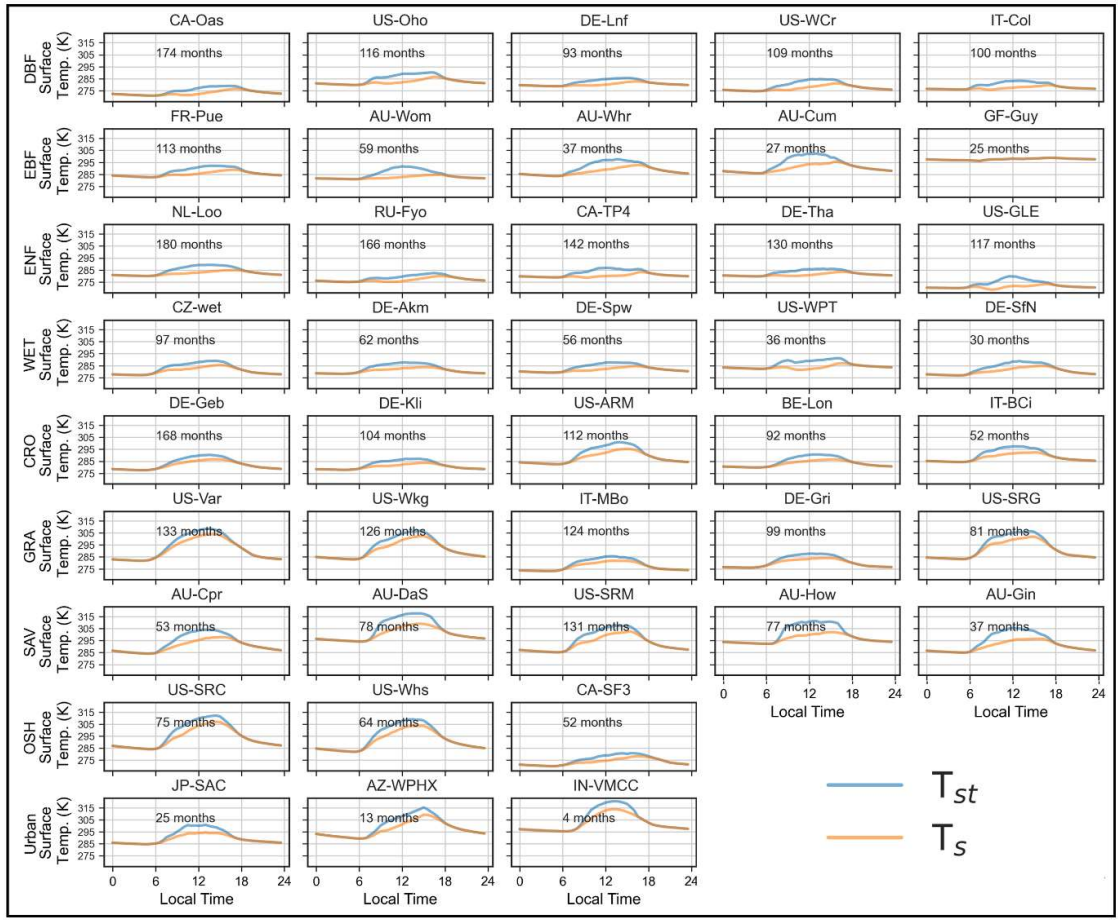

*Extended Figure 2: The diurnal averaged thermodynamically estimated Surface temperature ($T_{st}$) and the actual surface temperature measured from the outgoing longwave radiation ($T_s$) for all sites. The figure shows $T_{st}$ higher than $T_s$ at all the sites.*

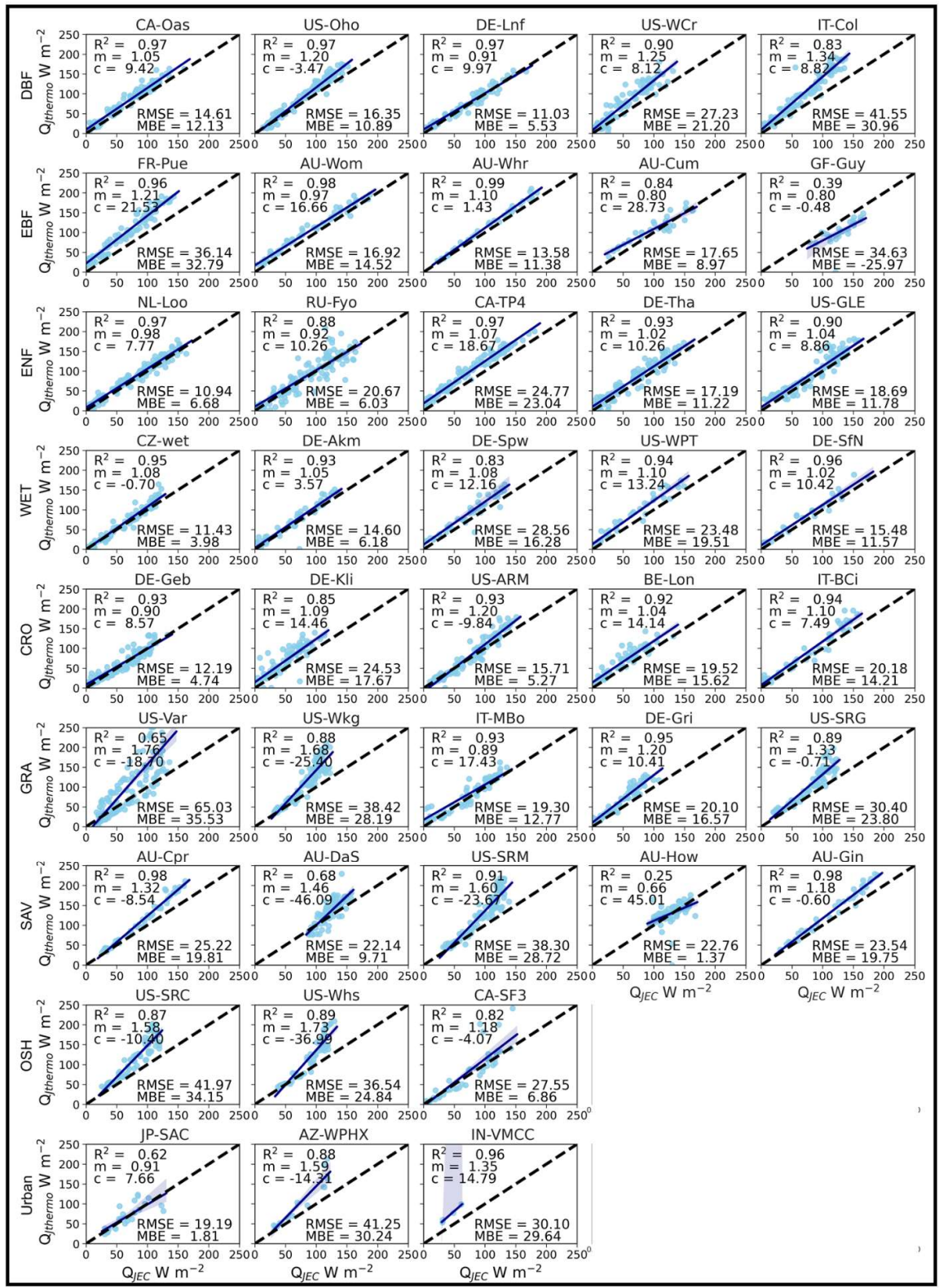

*Extended Figure 3. Evaluating the estimations $Q_{Jthermo}$ with $Q_{JEC}$ using correlation statistics. The data points are at the monthly resolution. The adjusted $R^2$ is the explained variance of $Q_{Jthermo}$ by $Q_{JEC}$ and a measure of the fit; slope (m); and intercept (c) are the regression coefficients. Root mean square error (RMSE) and Mean Bias error (MBE) are estimated for $Q_{Jthermo}$ with respect to $Q_{JEC}$.*

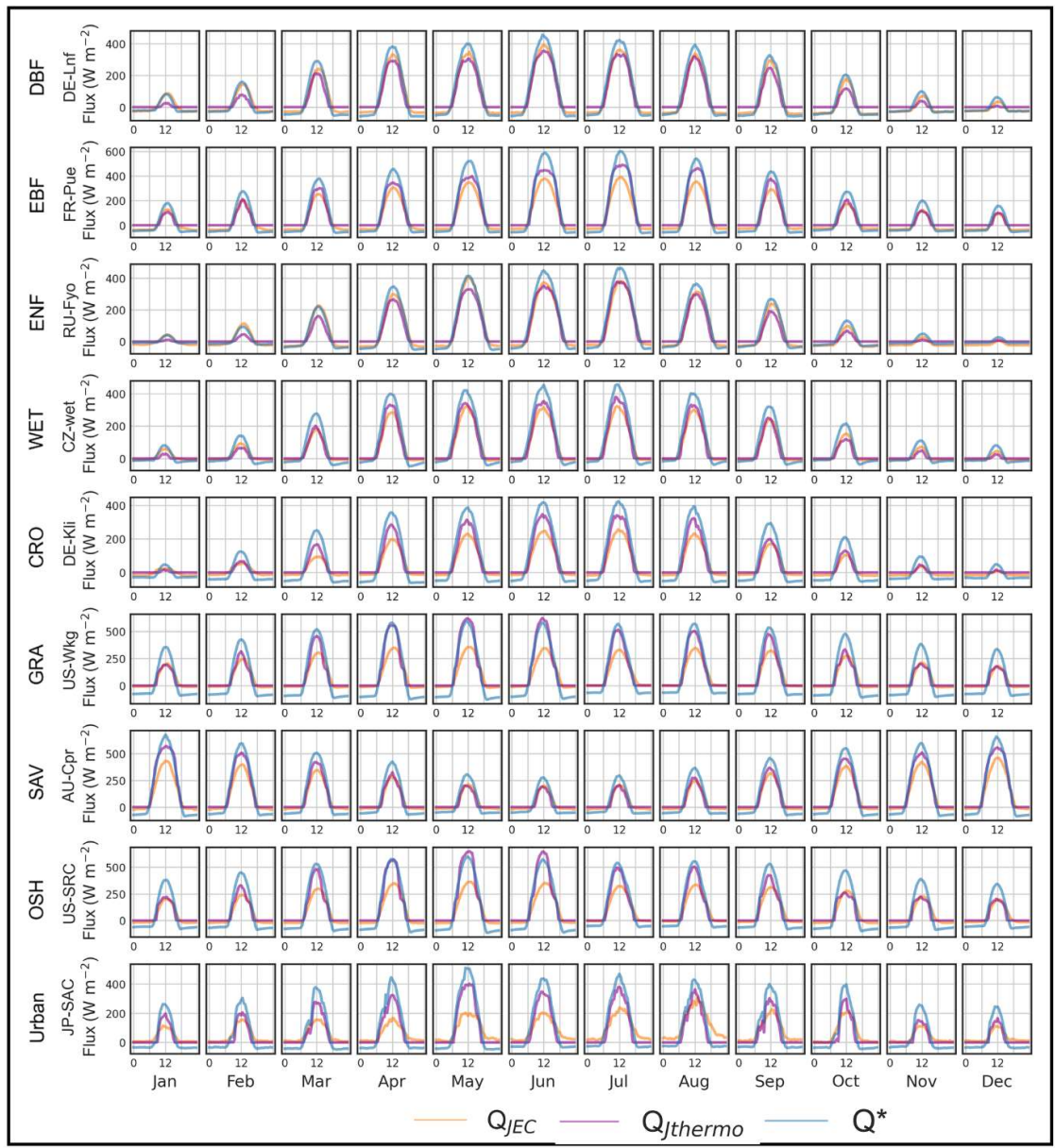

Extended Figure 4. The comparison between the monthly diurnal average fluxes from TE and EC observations for FLUXNET and urban sites based on different land covers. $Q_J$ (Observed turbulent heat flux by EC), $Q_{Jthermo}$ (turbulent heat flux estimated from thermodynamic model), and $Q^*$ (Observed net radiation).

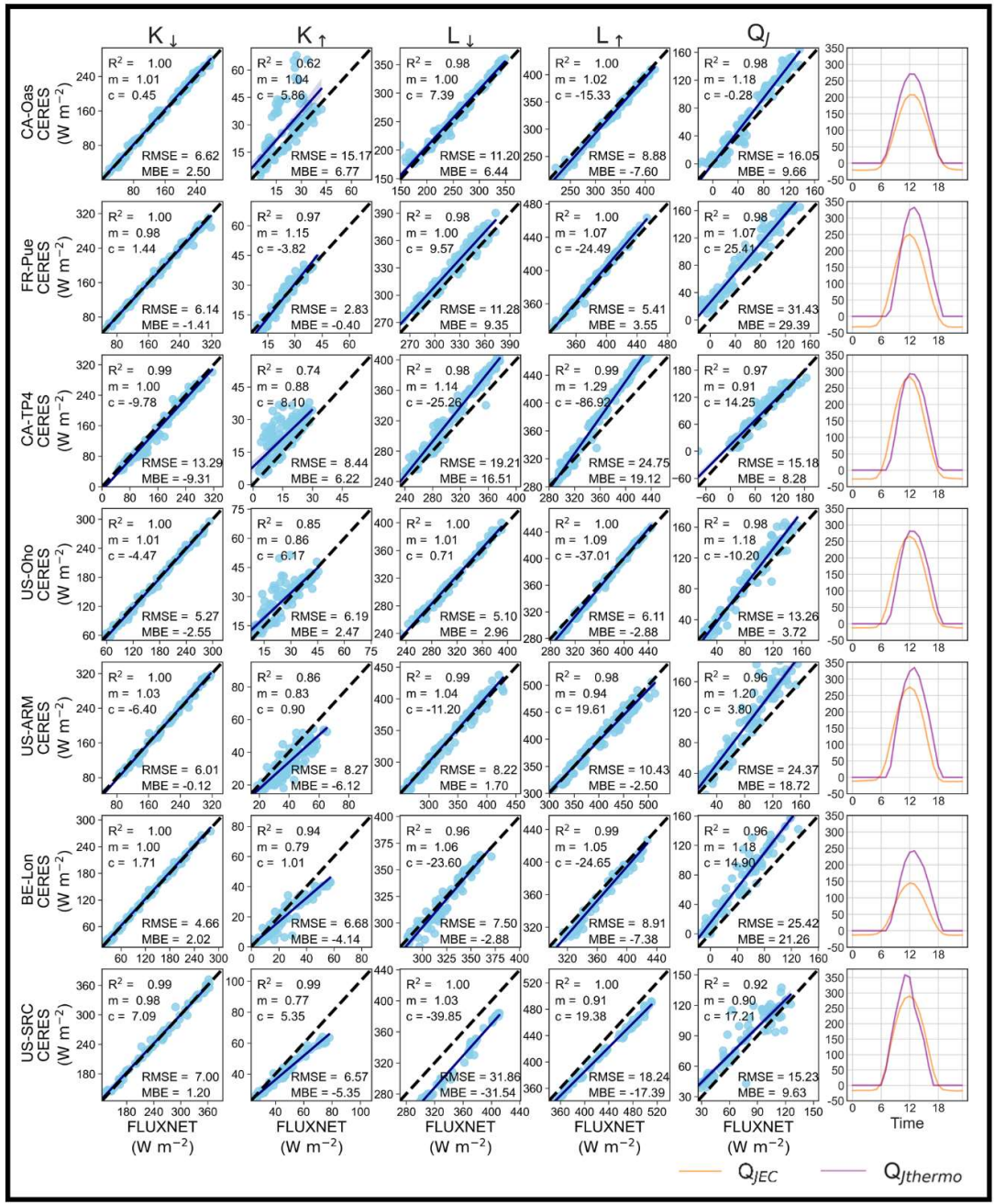

Extended Figure 5: Validation of monthly estimates of $Q_{Jthermo}$ (turbulent heat flux estimated from the thermodynamic model with inputs of radiation variables from CERES monthly global Syn dataset at 1° resolution) with $Q_{JEC}$ (Observed turbulent heat flux by EC from FLUXNET2015 databases). The figure also shows the variation of CERES Incoming shortwave radiation ($K_↓$), Outgoing shortwave radiation ($K_↑$), Incoming longwave radiation ($L_↓$), Outgoing longwave radiation ($L_↑$) with observed on-site data of corresponding variables on flux towers within the grid. The statistics used are very similar to Extended Figure 2.

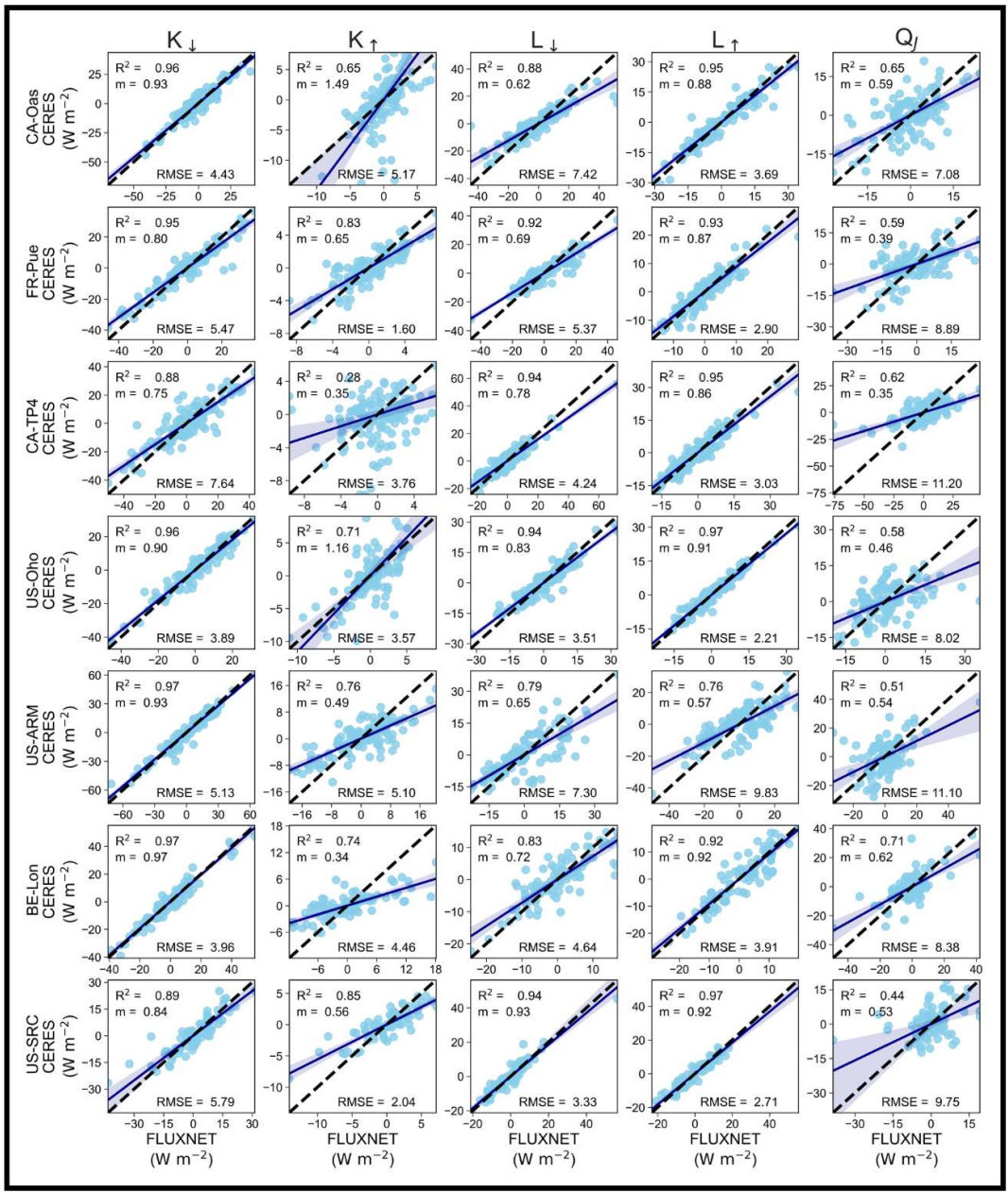

*Extended Figure 6: Variation of monthly anomalies between CERES and FLUXNET data estimated after deseasonalisation of all the variables.*

*Extended Tables:*

Extended Table 1 : Evaluation of $Q_{Jthermo}$ (turbulent heat flux estimated from the thermodynamic model) with $Q_{JEC}$ (Observed turbulent heat flux by EC): N (Number of Sites); RMSE (Root Mean Square error in Wm$^{-2}$), MBE (Mean Bias error in Wm$^{-2}$), $R^2$ (Variation explained by Correlation), m (Slope) and c (intercept) of regression line expressed as mean ± Standard Deviation.

| Land Cover | N | RMSE | MBE | $R^2$ | m | c |
|---|---|---|---|---|---|---|
| **All** | 102 | 23.2 ± 10.9 | 11.9 ± 13.1 | 0.86 ± 0.15 | 1.12 ± 0.27 | 1.9 ± 21 |
| DBF | 13 | 23.4 ± 9.5 | 17.2 ± 10.3 | 0.9 ± 0.15 | 1.05 ± 0.23 | 14.6 ± 17.2 |
| EBF | 8 | 24.5 ± 8.6 | 5.5 ± 19.7 | 0.74 ± 0.27 | 1.06 ± 0.29 | -1.4 ± 38.1 |
| ENF | 22 | 19 ± 8.5 | 9.1 ± 12.6 | 0.93 ± 0.05 | 1.02 ± 0.12 | 7.4 ± 12.4 |
| WET | 12 | 18.8 ± 6.7 | 9.4 ± 7.4 | 0.88 ± 0.1 | 1.05 ± 0.29 | 7.9 ± 22.5 |
| CRO | 9 | 21.4 ± 8.2 | 9.8 ± 10.1 | 0.87 ± 0.15 | 1.06 ± 0.24 | 3.9 ± 11.3 |
| GRA | 22 | 24 ± 13.4 | 10.7 ± 13.2 | 0.86 ± 0.11 | 1.21 ± 0.29 | -6.2 ± 17.3 |
| SAV | 10 | 30.6 ± 14.8 | 18.3 ± 15.4 | 0.72 ± 0.23 | 1.29 ± 0.3 | -10.2 ± 26.7 |
| OSH | 3 | 35.4 ± 7.3 | 21.9 ± 13.9 | 0.86 ± 0.04 | 1.5 ± 0.28 | -17.2 ± 17.5 |
| Urban | 3 | 30.2 ± 11 | 20.6 ± 16.2 | 0.82 ± 0.18 | 1.28 ± 0.34 | 2.7 ± 15.2 |
| **Forest** | 43 | 21.3 ± 9 | 10.8 ± 13.9 | 0.89 ± 0.16 | 1.04 ± 0.19 | 7.9 ± 20.8 |
| **Non-Forest** | 31 | 23.3 ± 12 | 10.5 ± 12.2 | 0.86 ± 0.12 | 1.17 ± 0.28 | -3.2 ± 16.3 |
| **Others** | 13 | 31.7 ± 13.3 | 19.1 ± 14.6 | 0.75 ± 0.21 | 1.34 ± 0.3 | -11.8 ± 24.4 |

Extended Table 2 : Analysis of bias between $Q_{Jthermo}$ (turbulent heat flux estimated from the thermodynamic model) with $Q_{JEC}$ (Observed turbulent heat flux by EC): N (Number of Sites); $Q_{JEC}/Q^*$ (EC Turbulent flux by Net radiation), $Q_{JEC}/Q^*$ (Thermodynamic estimate of Turbulent flux by Net radiation), and Bias/$Q^*$ (Bias between $Q_{Jthermo}$ and $Q_{JEC}$ by Net radiation) expressed as Percentage mean ± Standard Deviation.

| Land Cover | N | $Q_{JEC}/Q^*$ (x100)% | $Q_{Jthermo}/Q^*$(x100)% | Bias/$Q^*$(x100)% |
|---|---|---|---|---|
| **All** | 102 | 61.0 ± 31.1 | 72.6 ± 8.6 | 11.6 ± 26.5 |
| DBF | 13 | 66.9 ± 11.9 | 73.5 ± 5.4 | 6.5 ± 11.6 |
| EBF | 8 | 70 ± 11.1 | 74.7 ± 2.5 | 4.7 ± 10.8 |
| ENF | 22 | 71.2 ± 12.3 | 70.8 ± 5.6 | -0.4 ± 15.2 |
| WET | 12 | 23 ± 76.9 | 63.6 ± 15.1 | 40.6 ± 64.2 |
| CRO | 9 | 59.2 ± 10 | 71.6 ± 7.5 | 12.3 ± 7.3 |
| GRA | 22 | 64.1 ± 11.4 | 73.4 ± 7.4 | 9.3 ± 9.2 |
| SAV | 10 | 68.7 ± 7.1 | 81.8 ± 3.2 | 13.2 ± 8.4 |
| OSH | 3 | 63.8 ± 7.1 | 77.7 ± 5.7 | 13.9 ± 12.8 |
| Urban | 3 | 44.8 ± 7.6 | 75.2 ± 8.4 | 30.5 ± 1 |
| **Forest** | 43 | 69.7 ± 11.8 | 72.3 ± 5.3 | 2.6 ± 13.5 |
| **Non-Forest** | 31 | 62.7 ± 11.1 | 72.8 ± 7.3 | 10.1 ± 8.7 |
| **Others** | 13 | 67.5 ± 7.1 | 80.9 ± 4.1 | 13.3 ± 8.9 |

Extended Table 3: Global annual land surface heat fluxes (mean ± SD Wm$^{-2}$) for the period 2003-2019 of different land use land cover estimated as area-weighted spatially averaged on IGBP global land covers from MODIS. The 2019 LULC was used for the analysis. The standard deviation here represents the interannual standard deviation.

| Land Cover | $Q^*$ | $H$ | $LE$ | $\Delta Q_s$ |
|---|---|---|---|---|
| **Global (90N-90S)** | **84.1 ± 0.81** | **42.4 ± 0.35** | **40.1 ± 0.48** | **2.21 ± 0.65** |
| Forests[1] | 112±1.04 | 33.72±0.35 | 71.5 ±0.48 | 6.72±0.6 |
| Non-Forest[2] | 93.9±1.04 | 48±0.62 | 43.4±0.62 | 2.7±0.88 |
| Others[3] | 91.7±0.9 | 45.1±0.76 | 45.4±0.95 | 1.5±0.73 |
| Barren | 71.18±1.35 | 65.35±0.8 | 8.55±0.4 | -2.47±1.0 |

[1]Forests comprises Evergreen Needleleaf Forests, Evergreen Broadleaf forests, Deciduous Needleleaf forests, Deciduous Broadleaf forests and Mixed Forests categories of MODIS land covers.

[2]Non Forest comprises Grasslands, Croplands and natural vegetation mosaics

[3]Others comprises Closed and Open Shrublands, Woddy Savannas and Savannas